\documentclass[11pt,a4paper]{article}
\usepackage[hyperref]{acl2020}
\usepackage{times}
\usepackage{latexsym}

\usepackage[colorinlistoftodos]{todonotes}

\usepackage{microtype}

\aclfinalcopy 

\title{``EHLO WORLD" - Checking If Your Conversational AI Knows Right from Wrong}

\author{Elayne Ruane \\
  School of Computer Science\\
  University College Dublin\\
  \texttt{elayne.ruane@ucdconnect.ie} \\\And
  Vivek Nallur \\
  School of Computer Science\\
  University College Dublin\\
  \texttt{vivek.nallur@ucd.ie} \\}

\date{}

\begin{document}
\maketitle
\begin{abstract}
In this paper we discuss approaches to evaluating and validating the ethical claims of a Conversational AI system. We outline considerations around both a top-down regulatory approach and bottom-up processes. We describe the ethical basis for each approach and propose a hybrid which we demonstrate by taking the case of a customer service chatbot as an example. We speculate on the kinds of top-down and bottom-up processes that would need to exist for a hybrid framework to successfully function as both an enabler as well as a shepherd among multiple use-cases and multiple competing AI solutions.

\end{abstract}

\section{Introduction}
\label{section:intro}
If someone claimed their AI system was ethical, in the sense that it operated according to some ethical principles, would you believe them? How should companies, individuals, and society communicate and continually validate claims about so-called ethical AI? 

There are many unanswered questions around how to design artificial moral agents if such a thing is even possible. The challenges and limitations of both top-down rule-based approaches and bottom-up iterative learning processes have been examined in the literature. In this paper we build on this work and discuss such approaches as they may apply specifically to Conversational AI, contextualised in the increased pervasiveness of this technology and in light of new regulations such as GDPR.

In most real-world deployments, explanations of each decision made by an AI system would be beyond the validating capacity of individuals. What is needed is the AI-behavioural equivalent of a hardware standard. That is, every company must ensure that any AI deployed by them should meet a minimum standard of behaviour. How do we design such a standard? The standard clearly needs to be implementation-agnostic. This paper speculates on whether top-down regulations, bottom-up processes or a combination of the two would best implement such a standard.

\section{Motivation}

There are several reasons why the ethics of Conversational AI is worth attention, not least of which is the increasing ubiquity of these agents in the last few years in particular. Since 2016 was named the ``Year of the Bot'' by Microsoft CEO, Satya Nadella \cite{reynolds2017satya_nadella}, chatbots and other Conversational Agents have been increasingly deployed in various domains including healthcare, education, customer service, personal organization, and social outreach. Recent work by \citet{ruane2019conversational} highlights ethical considerations around the social impact of Conversational Agents, and calls for a socially mindful approach to building Conversational Agents throughout the entire design and development process. The authors argue that as the social capital of Conversational Agents increases, so too does the responsibility on agent owners to more carefully consider how they design their agents, and the impact it can have on individuals, especially marginalized groups and society's most vulnerable.

Conversational Agents seek to mimic a very human behaviour - communication through nuanced language. As such, when we discuss ethical standards of behaviour for these agents it makes sense to look towards the study of normative ethics, i.e., how one \textit{should} act. The major schools of normative ethics that have emerged from moral philosophy aim to define good and evil according to a specific paradigm. Within each school there are a number of moral theories that we can use to define actions as either right or wrong according to the theory's principles in context of the school of thought. But how can we apply these concepts to Conversational Agents?

Conversational Agents as moral agents have received less attention in the literature compared to general AI-systems. At what point does the Conversational Agent become a moral agent? An agent that cannot reason but simply carries out dialogue as designed by the agent owner cannot be said to be ``responsible’’ for its ``actions’’, but the designer is responsible. When a Conversational Agent can learn beyond the scope of explicit dialogue defined by a human, does it become a moral agent? There is overlap here between responsibility and morality. Ultimately, humans who build Conversational Agents are responsible for their social impact, whether they can predict and control how the system works or not. However, unlike how the concept of a company as an artificial person has legal status, AI systems including Conversational Agents lack clear legal status and definitions of responsibility for agent owners. As such, when we discuss morality of these agents through actions they should and should be designed not to take, we do so without legal structure. Many philosophical and moral theories and positions are based on a spiritual world view. Humans are presumed to live for themselves, and have innate value and rights by virtue of their consciousness. Unlike humans, AI systems are usually built for specific goals which they aim to achieve on behalf of someone or some organisation. As such, when we discuss machine ethics, we cannot separate the machine from the goals of the owner. We may need a new view of ethics that does not place the motivation or onus on the inherent ``moral sense'' of the agent.

\section{Background}

\begin{figure}[h!]
  \caption{Chatbot Architecture from V-Soft Consulting\footnote{Source: https://blog.vsoftconsulting.com/blog/understanding-the-architecture-of-conversational-chatbot}}
  \includegraphics[width=\linewidth]{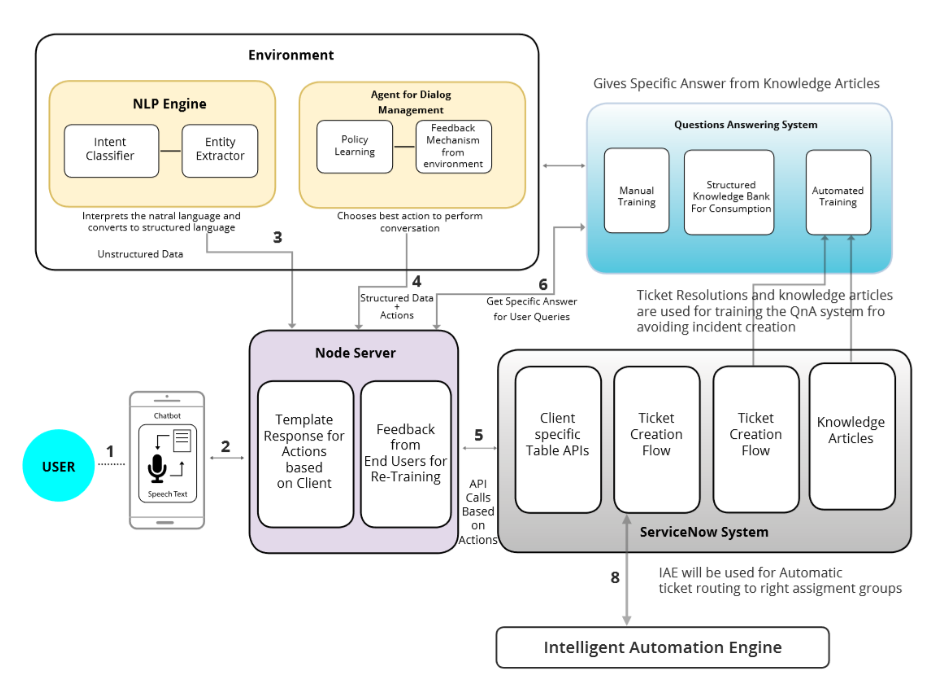}
  \label{fig:chatbot_architecture}
\end{figure}

Just as there are two sides to the conversation, there are also two sides to the development of conversational AI; natural language understanding (NLU) and natural language generation (NLG), see Figure \ref{fig:chatbot_architecture} for detailed example architecture. NLU is required to interpret the user’s utterances and determine their meaning and intent. NLG is required to generate the agent’s responses and may be based on some specific response strategy. Models developed to understand user input usually employ supervised or semi-supervised learning strategies that leverage labelled datasets of conversational utterances. These may be from conversation logs, or other domain-specific sources. Once the user intent is identified and some meaning has been derived from the utterance, the agent needs to craft a response. Many chatbots use a rule-based approach for this which amounts to a number of if-else statements: if the user intent is X, respond with Y. Depending on how much effort is dedicated to crafting responses, there may be multiple available responses out of which a random selection or some criteria-based selection occurs. It is clear to see how the dialogue designers’ own values could be encoded into the design of the agent in this case. The advantages of this method are control over the quality of agent responses for all defined intents. Of course, this approach becomes less viable as the scope of the conversation increases, and can lead to an inflexible and repetitive conversational experience for users. Despite that, it is a common approach, particularly where large training data sets or computational resources are unavailable. Automatic response generation may be implemented using supervised or unsupervised techniques. Without significant amounts of training data and much, often manual, evaluation, this approach will produce poor quality dialogue and may risk harming user experience. In this case, the ``values’’ that are baked into the system will reflect any bias in the training dataset. Designing agent dialogue is one aspect of the process. Other considerations during development include how to ask the user for informed consent in relation to data collection, how to record and store conversation logs including personally identifiable information, how to design and represent agent persona, and how to evaluate user experience and protect users from harmful replies.

\section{Ethics of Conversational AI}

More advanced Conversational Agents such as the personal assistant offering from major tech companies like Apple's Siri, Google Assistant, Amazon Alexa, and Microsoft's Cortana allow the user to interact conversationally while completing a range of tasks such as using a search engine, screening phone calls, playing media, getting restaurant recommendations, getting weather and news reports, booking appointments, and setting reminders. The list of supported tasks is long and when a user engages in these tasks they quickly generate a descriptive user profile. Not only is the agent able to gather personally identifiable information (PII) but also additional data such as users' likes and dislikes, their daily routine, social connections, and future plans. Users are encouraged to self-disclose to improve their experience through customization. Users stand to gain in the short-term, when they disclose information \cite{saffarizadeh2017conversational}. Additionally, users often perceive their interaction to be ``anonymous'', or at least isolated to their own device, and this perceived anonymity can encourage self-disclosure \cite{evans2010impact}. Recent work by \citet{xiao2019tell} found when a survey was conducted via a Conversational Agent, users were more engaged and provided more informative answers compared to a traditional survey form (n=600). This highlights the impact of conversation on user engagement and thus self-disclosure. 

As it should, discussions of self-disclosure centre around user privacy, an important and increasingly examined topic in  literature. However, there is another dimension to consider: user manipulation. When an agent is customer-facing and representing an organisation, a clear question arises: whose interest should it serve? This is an ethical consideration as old as commerce itself; does one put the customers' needs over the company's bottom line? This is a well-known tension. In the case of a human sales representative, especially one who works on commission, there is often an expectation that they may try to up-sell the customer. The company expects them to meet user needs and thus make sales, but the customer \textit{knows} that the salesperson likely puts their own, and possibly the company's, interests first. They are expecting to be \textit{sold to}. However, when a user is dealing with an automatic agent such as a product recommendation chatbot, they may not have the same intuition. The user may expect \textit{information} about the products and assume it to be largely unbiased. The key difference here is awareness on the customer's part of how the agent was designed. From a business perspective, the design approach will likely depend on the nature of the sale. But should this behaviour be governed by ethical principles, or the company's profit margins? 

We already know companies try to employ dynamic pricing to maximise profits. Other techniques designed to encourage purchases include pressure tactics such as those employed on Booking.com (and other sites) that show notifications such as ``x number of people are looking at this right now'' to try and manipulate the user into purchasing quickly. However, users discovered that these numbers were generated by a JavaScript call to \texttt{Math.random()}! Refreshing the page would generate a new number of people supposedly viewing the property. Such tactics, including but not limited to, dynamic pricing could be even more effective and less transparent in the context of Conversational AI because the user can be manipulated using the large amount of personal information they have already disclosed. Based on the social impact of these agents and the potential to harm users, there is a strong case to investigate how ethical concerns should be addressed and, subsequently, how claims of ethical standards or moral behaviour should be validated.

\section{Related Work}


\subsection{Ethics in AI}

Most current research on ethics in AI seems to focus on two key strands: Explainable AI and Bias in AI. Explainable AI is concerned with shedding light on how an AI system arrives at a particular decision. This is based on the idea that if we can understand the reasoning process, we can evaluate whether it was ethical or not. On the surface this appears to be based in virtue ethics, where we focus on the agent's ``intent'', i.e. what features did the system use to make this decision? For some systems this approach makes sense; we cannot take a consequentialist approach because we may not be able to evaluate the impact of an agent's decisions due to limitations in computational resources or access to necessary information. Given the lack of consensus in philosophy about the `correct'`’ set of ethics (for a human or even a machine), it is difficult to recommend a deontological approach. Much attention is therefore given to the problem of bias in training data and how this can affect the predictions or classifications made by a system. However, for most commercial implementations, training datasets hold significant value and are protected as trade secrets. Additionally, data protection laws such as GDPR that protect consumers may mean this data cannot be easily shared for third-party validation/verification. 

\subsection{Machine Ethics}
In the introduction to the special issue of Ethics and Information Technology, \citet{dignum2018ethics} presents a number of open questions around the moral, social, and legal consequences of decisions made by AI systems. Dignum calls for frameworks to guide design choices, and for theories and methods to evaluate decisions at all stages of development. An important argument highlighted here is the need to make explicit, the implicit values held by various stakeholders that are further dependent on their socio-cultural context. Dignum argues an autonomous AI systems should be a responsible in so far as it should take into account societal values, moral and ethical considerations, and weigh the priorities of different stakeholders. 



\citet{rahwan2018SITL} proposes a society-in-the-loop approach which takes human-in-the-loop ideas from Machine Learning for expert, uncontested tasks and combines it with social contract theory in order to embed societal values into algorithmic governance for tasks or decisions where stakeholders may have conflicting interests and values.  The current AI landscape could be considered to be in a semi-State of Nature, the starting point for social contract theory as described by \citet{hobbes1967leviathan}. Currently AI systems are free to act as designed because we lack the appropriate ethical principles to apply to them or ways of reliably evaluating their behaviour. However, we also do not currently have technology that is capable of autonomously deciding to: kill humans, disobey orders, manipulate its own code, or refuse to be shut down. In social contract theory, morality is derived from a set of rules that rational individuals agree to obey for the mutual benefit of social order. This is what \citet{rahwan2018SITL} argues for while acknowledging that a key limitation of this approach is the difficulty in articulating social values and the limits of public engagement on nuanced issues. In a similar vein, \citet{rawls19711999} presented a thought experiment to identify moral acts as those everyone would agree to if they were unbiased. He describes a thought experiment in which individuals striving to cooperate act behind a ``veil of ignorance'', blind to their gender, race, age, intelligence, wealth, skills, education, and religion. Principles of justice formed in this unbiased way could be considered to be fair. This may be an approach that could be used to develop an idea of fairness in Conversational AI where stakeholders have conflicting interests. If the veil of ignorance masked whether the participant was an agent owner, user, or some other stakeholder, would they consider a design decision to be fair? This is an example of a top-down approach where morality is defined or described, and the agent is developed accordingly.

\citet{arnold2018redbutton} discuss the limitations of a ``big red button'' approach - i.e. when a system starts making decisions outside acceptable bounds, a stop-measure could be used to shut it down. The authors argue this option would already be too late because some as-yet undefined harm would have to occur to warrant shutdown. The authors also detail concerns such as if the system somehow got control of and disabled its own shutdown button e.g. maximising its reward function by disabling system shutdown, or if a human couldn't press the button in time to avoid further harm. Instead, the authors argue for an ethical core (EC) component that generates scenarios and evaluates the system's decisions in an on-going manner. They argue this is in line with required standards of safety; we don't want to simply avoid disaster, we want reliable and consistent safe, legal, and moral behaviour. This work is an example of algorithms governing algorithms but it is not clear how the ethical core would establish whether the system was operating outside of ethically acceptable behaviour because much of the focus of the paper is on ensuring that the system cannot determine it is being tested and then manipulate the test by altering its behaviour.


\subsection{Artificial Morality}

Possibly the most closely related work and on which this paper builds is that of \citet{allen2005artificial} who discuss how artificial morality can be implemented, that is, designing agents that have a sense of ethics and moral reasoning. The authors highlight that AI systems take part in activities that have moral consequences and thus their actions should be evaluated. If this evaluation is not possible for humans to do due to complexity or speed of computation, then it follows that systems should have some ethical and legal understanding ``built in'' such that they can be monitored by themselves or by another automatic process. Some systems can be treated as ethical black boxes, and the output evaluated, however in other scenarios the authors argue ethical reasoning should be considered an engineering design challenge that tackles difficult questions such as how to make explicit ethical decisions. This work was published in 2005 and we've now seen where such challenges have come to light in the context of autonomous vehicles who must have decision-making abilities embedded into their algorithms that are deeply ethical in nature such as who should the car prioritise in a safety-critical scenario, a passenger or a pedestrian? 
Allen et al. look at top-down approaches that aim to turn explicit theories of moral behaviour into algorithms that can be used to develop artificial moral agents (AMAs). Top-down approaches suffer from a limitation of any rule-based approach to intelligent behaviour; they don't generalise well to real world complex scenarios. Such a deontological approach requires a set of ethical or moral rules to implement. The authors highlight some well-known sources including religious rules such as the Ten Commandments but note that selecting guidelines from any such source would naturally be controversial. The authors note a strength of this approach is the ability to add rules for specific situations. This is particularly effective where the system's scope is both limited and well defined. However, a major limitation is the conflict rules can cause when they require different actions in the same scenario and these conflicts cannot be computationally resolved. The authors look towards moral philosophy for guidance including consequentialist and deontology ethics. The authors note utilitarianism's Achilles heel is the measurement problem, and would also require significant computation to evaluate the consequences of alternative actions even if some measurement criteria could be agreed upon. The authors turn to Kant's categorical imperative as a higher principle that would not require evaluating specific actions that may conflict. The authors discuss the computational burden on such a system and the level of understanding of human psychology and how actions could affect all humans. Although these approaches are limited by computational capacity and availability of information, human beings are capable of acting morally within similar constraints. A key difference is the idea of phronesis, or lived experience. 
The work also looks at bottom-up approaches in an attempt to train an agent to mimic moral human behaviour (phronesis). Bottom-up approaches do not require a set of rules that should be followed but instead try to provide scenarios in which the ``right'' or moral action is rewarded. This may be implemented using reinforcement learning. This approach is akin to how humans learn to behave morally, through interaction. While top-down approaches are difficult to conceptualize and make explicit, bottom-up processes are difficult to implement. The authors raise concerns that have been echoed in later works such as \citet{arnold2018redbutton} discussion of an ``Ethical Core'', around how a system that has the capacity to learn ethical behaviour can also learn unethical behaviour.

The authors suggest some kind of blind test that would compare a system's performance to that of a human on a set of moral judgements tests like a Moral Turing Test \cite{allen2000prolegomena}. Of course, this test would suffer from the same limitations of the Turing Test in trying to evaluate intelligence. 


\section{Top-down Regulation}
\label{section:reg}

There are many vision papers about top-down regulation and what it would look like. Previous discussion around user manipulation in Section \ref{section:intro} highlights how specific legal requirements need to be met in order to ensure ethical design. GDPR is an example of legislation that explicitly addresses data privacy and consent issues and requires transparency in relation to these concerns. However, it is not AI-specific and while it addresses AI-related concerns such as decisions about individuals as a result of automated decision making and profiling\footnote{EU GDPR Article 22: https://www.privacy-regulation.eu/en/article-22-automated-individual-decision-making-including-profiling-GDPR.htm}, these requirements apply to a narrow set of decisions, and the subsequent requirements leave some room for debate. 

This ambiguity can be associated with the general difficulty in creating guidelines for AI. \citet{jobin2019global} showed companies, research institutions, and public sector organizations around the world have all issued guidelines for ethical artificial intelligence and there appears to be global convergence around five ethical principles: transparency, justice and fairness, non-maleficence, responsibility, and privacy. However, the authors discovered substantive divergence in how these are interpreted, determined significant, in which scenarios they apply, and the recommended implementation strategies, noting that there was no single principle that appeared in all of the 84 guidelines they reviewed! 

From an interaction perspective, this ambiguity allows companies to use specific language to remove agency from the user. We can see how GDPR guidelines around consent and cookies has been implemented across various websites which use legal jargon to obfuscate cookie preferences such that the user will simply accept recommended cookies (including advertising). Similarly, a chatbot that uses ambiguous language when asking for user consent to process their data is taking advantage of limitations of the user's understanding of the system and what can be done with their data. While GDPR is a good starting point, AI systems and even Conversational AI need their own specific regulations and even still, these are unlikely to be able to guarantee ethical implementation on their own. Another challenge for a top-down regulation approach is that it is, by nature, a slow process and it cannot seem to keep pace with technological development because the pace of innovation is much faster than the deliberation of complex social and ethical questions, and the drafting of legislation or guidelines to address them.

How could we validate that an AI system is adhering to set out regulations? One option is a simulation engine, much like the ``ethical core'' proposed by \citet{arnold2018redbutton}, whereby the system is continually tested in simulated scenarios that require it to make decisions, where multiple good, bad, and ambiguous choices are available to it.Although it remains to be seen how computationally expensive this option would be. A system could be said to meet expectations if it reliably makes good choices in these situations. A challenge with this approach is the difficulty of explicitly coding these ethical principles into the system and designing scenarios to test them. This process should be led by the guidelines and regulations that are being evaluated and highlights the need for explicit and AI-specific regulations. 


The experience of GDPR and how agent owner values impact design of an agent shows us that the best any top-down regulation can achieve is to emphasize the achievement of some abstract value. \citet{hutchinson201950Years} illustrates quite clearly, how these abstract values are themselves contentious and consequently the yardsticks and metrics used to measure the achievement of said value may be of little use. This leaves us with the notion of designing-in or modifying the bottom-up processes that a company uses to create Conversational Agents, to achieve our desired goal. This has been done previously in the area of software testing. The effectiveness of automated unit-tests, integration tests and continuous build processes has led to large software projects being successfully developed and deployed.


\section{Bottom-up Processes}

By a bottom-up process, we mean the introduction of either an activity, artefact, or a combination of the two, such that a particular team-defined unit of code progresses to the next stage of the team's development/build/deployment cycle. The bottom-up process may certify that the unit of code, either sufficiently meets pre-defined criteria or possesses pre-defined attributes according to team policy. The process need not be a boolean-style (``test pass / test fail'') gatekeeper, but rather it could result in any measure that can be compared with another unit of code's performance on that process.

\paragraph{Bias Removal Checklist}
As mentioned previously, opening up datasets to public scrutiny might be difficult due to privacy concerns and release of technological secrets. However, regardless of the datasets being used, it is possible to have open-to-scrutiny \textit{bias removal checklists}, which contain statistical tests to be run and requirements for measures to be publicly recorded. For any AI library that is used by a company, the library must be certified to have cleared the bias removal checklist. 

\paragraph{Open-Source Certified Components}
Take the scenario of a customer service Conversational Agent which requires personally identifiable information (PII) to validate the user and subsequently address their query. Customer interactions need to be logged for a number of legal and operational reasons, but PII should not be stored in plain text logs that have team-wide or department-wide access. A possible solution is partially encrypted logs or distributed logs. This requires reliably identifying PII which may not be difficult when this information is provided by the user in response to a question like ``What is your account number?'' but becomes more difficult when the user inadvertently or otherwise self-discloses sensitive information. How can we validate that the logs are encrypted? An open source/certified persistence layer that encrypts PII can be devised.  Repeatable builds~\cite{heydon2001vesta} are a software engineering feature that every professional software organization strives to create through its build process. Up until now, these repeatable builds were useful for the software developer and maintainer, to find and reproduce bugs. However, now these can be used by a conformance tester to evaluate whether the deployed software used a particular unbiased dataset or behaviour-selection procedures or not. Using mechanisms such as build verifiability, we can approach the equivalence of certification. That is, for any given build of a software, there must be a repeatable build process that always uses the open-source/certified persistence layer and automated test cases check whether this has been done. The automated test cases' status would act as a certifying label for the deployed  conversational agent. A chatbot could disclose its certification status to the user to build trust. We already accept such tests from certification authorities in the field of electronics manufacture, medical devices, etc. 

This would seem to suggest that the presence of bottom-up processes would be sufficient to ensure that only trustable code is being deployed. However, the intent of the bias removal checklist, as well as automated certification of continuously deployed software all depend on 'governance credentials' of the checklist creator as well as the certification test creators. These credentials could develop in a self-organizing manner by a voluntary body. However, a more efficient process that society has settled on for other domains, such as pharmaceuticals, is the presence of a neutral third-party, established by regulation.

\section{Hybrid Process}

We discussed in Section \ref{section:reg} how regulations that are open to interpretation make it easy to circumvent a top-down process. The lack of a bottom-up process that can also guarantee ethical behaviour is a disappointment. That is why we argue for a combination of both. Regulation largely focuses on what an agent \textit{should NOT do} whereas a certificate can focus on what an agent \textit{will do} in an ethical sense. It is unlikely that regulation will cover every case, and context, where conversational AI could be gainfully employed. This would mean that regulation would have to leave space for bottom-up, decentralized processes to flexibly adapt to various situations. However, the impact of a rogue company or developer given free rein to make their own judgements could be quite harmful to vulnerable groups in society. This implies that, in reality, there would need to be an inter-dependent framework of top-down regulation as well as bottom-up processes, to ensure that conversational AI is only used in an ethical manner. We can see how this is mirrored in both safety legislation and certifications. It is already accepted practice that if a product is defective it is recalled, destroyed, and possibly the company responsible is fined. The company then has to fix its bottom-up processes to avoid the same malfunction in the future. 

\begin{figure}[h!]
  \caption{Aggressive Selling Techniques}
  \includegraphics[width=\linewidth]{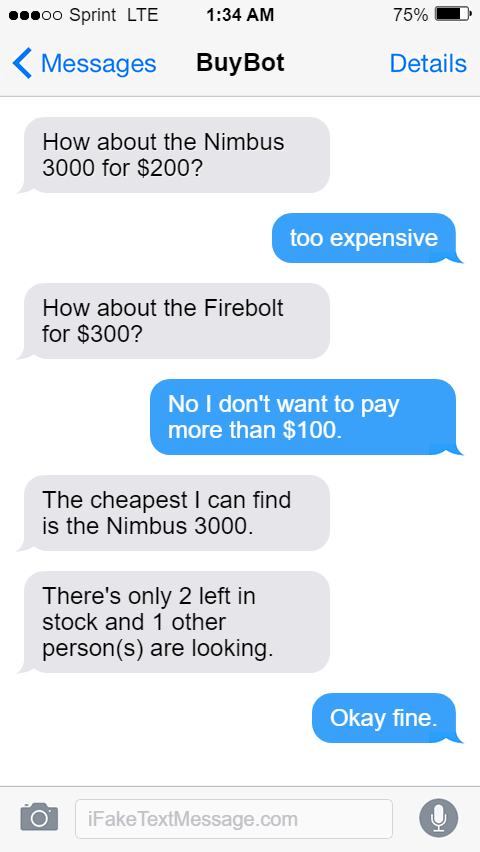}
  \label{fig:bully_bot}
\end{figure}

While general ethical concerns such as transparency, privacy, trust, and social responsibility around Conversational Agents are true for most or all conversational systems, the specific ethical issues and considerations that arise vary markedly depending on the design of the agent, its use case, and the profile of the user group \cite{ruane2019conversational}. Thus, a top-down regulation for conversational agents might specify that a chatbot must in no case \textit{bully} a user into making a decision (see Figure \ref{fig:bully_bot} for an example where the user does not have a way to validate the agent's stock claims). The bottom-up process might include a checklist that specifies the maximum number of times a customer's negative intent-to-buy is countered with an offer. In another example, a top-down regulation might specify that in no case should PII that is revealed during a conversation be stored. The bottom-up process to meet this regulation could be build-verifiability of using a particular version of a certified component, along with test cases that attempt to search for any PII in stored conversation logs.

\section{Conclusion}

It is clear that while there are many questions to be addressed around ethical conversational AI, a hybrid approach that leverages both top-down regulation to govern what a Conversational Agent should not do, and bottom-up processes that validate what it will do is preferable to relying on either approach in isolation. It is not the contention of this paper, that one  particular top-down regulation or one particular bottom-up process is the magic sauce that makes a chatbot ethical. Rather that we will need a set of inter-dependent processes that will incentivize small, but verifiable claims about conversational agent behaviour and agent-owner ethics.

\bibliography{acl2020}
\bibliographystyle{acl_natbib}

\end{document}